\documentclass[11pt]{article}
\usepackage{amssymb,graphicx,epsfig,cite}
\begin{document}
\thispagestyle{empty}

\hfill TIFR-TH/14-15

\bigskip

\begin{center}
{\LARGE\bf Anomalous Triple Gauge Vertices at the \\ [2mm]
Large Hadron-Electron Collider}

\bigskip

{\large\sl Sudhansu S. Biswal}\,$^{a,1}$,
{\large\sl Monalisa Patra}\,$^{b,2}$
and
{\large\sl Sreerup Raychaudhuri}\,$^{b,3}$
 
\bigskip 
 
{\small
$^a$ College of Basic Sciences, Orissa University of Agriculture and 
Technology, \\ Bhubaneshwar 751~003, India. \\ [2.5mm]
$^b$ Department of Theoretical Physics, Tata Institute of Fundamental 
Research, \\ 1 Homi Bhabha Road, Mumbai 400005, India.
}
\end{center}

\bigskip\bigskip

\begin{center} {\Large\bf Abstract} \end{center}
\vspace*{-0.35in}
\begin{quotation}
\noindent 
At a high energy $ep$ collider, such as the Large Hadron-Electron 
Collider (LHeC) which is being planned at CERN, one can access the 
$WW\gamma$ vertex exclusively in charged current events with a radiated 
photon, with no interference from the $WWZ$ vertex. We find that the 
azimuthal angle between the jet and the missing momentum in each charged 
current event is a sensitive probe of anomalous $WW\gamma$ couplings, and 
show that for quite reasonable values of integrated luminosity, the LHeC 
can extend the discovery reach for these couplings beyond all present 
experimental bounds.
\end{quotation}

\bigskip

\centerline{{\sf Pacs Nos:} 14.70.Fm, 12.60.-i, 12.15.-y, 12.20.Fv}

\vfill

\centerline{\today}

\bigskip

\hrule
\vspace*{-0.1in}
$^1$ sudhansu.biswal@gmail.com \hspace*{0.35in} 
$^2$ monalisa@theory.tifr.res.in \hfill 
$^3$ sreerup@theory.tifr.res.in

\newpage

The Standard Model (SM) of elementary particle physics, originally 
proposed\cite{SM} in the 1960's, has achieved completion with the 
near-certain discovery in 2012\cite{HiggsDiscovery} of the long-predicted 
Higgs boson \cite{Higgs}. This became possible only because of the 
commissioning of the Large Hadron Collider (LHC) at CERN, Geneva, a high 
energy machine which runs with a greater collision energy than any of its 
predecessors could achieve. The LHC is currently shut down for 
significant upgrades in energy and luminosity intended for its next run 
in 2015. In the community of high energy physicists there are high 
expectations that in that run, or in following years, the LHC might 
conclusively find some signals that the Standard Model of particle 
physics is not the final theory, but simply an effective theory which has 
worked efficiently to explain the experimental results collected till 
date, but which will prove inadequate when we go to higher energies. In 
this article, we do not plan to go into the multiple reasons for such an 
expectation, which are well-discussed in the literature\cite{BSM}, but 
instead focus on one of the possible ways in which such signals for new 
physics beyond the SM could be found.

\begin{figure}[h!] 
\begin{minipage}[h]{0.47\linewidth}
\centerline{\epsfig{file=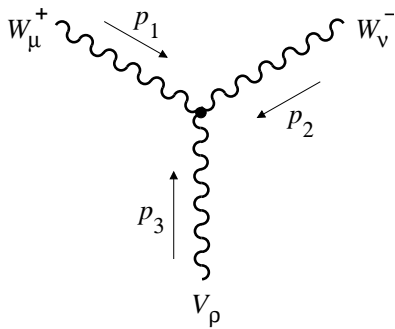,scale = 0.35}}
\vspace*{-0.1in}
\caption{\footnotesize Illustrating momentum assignments for the generic 
$WWV$ vertex.}
\label{fig:vertex}
\end{minipage} 
\hspace{0.02\linewidth}
\begin{minipage}[h]{0.51\linewidth}
The specific part of the SM on which we focus is one of the {\it triple 
gauge boson vertices} (TGV's) in the Standard Model --- more 
specifically, the $W^+W^-V$ vertex. Here $V$ can denote any one of the 
neutral vector bosons $\gamma$ or $Z$, but in this work, we focus on the 
specific case $V = \gamma$. In the Standard Model, of course, this vertex 
is precisely defined\cite{TGVSM}. However, it is also possible to 
parametrise possible new physics contributions to this 
vertex\cite{anomalous} in the form of a pair of undetermined parameters 
$(\Delta\kappa_\gamma,\lambda_\gamma)$.
\end{minipage} 
\end{figure}
If we denote the $W^+_\mu(p_1)W^-_\nu(p_2)A_\rho(p_3)$ vertex by
$i\Gamma^{(WW\gamma)}_{\mu\nu\rho}(p_1,p_2,p_3)$,
then it can be neatly parametrised in the form of three separate terms, viz.
\begin{equation}
i\Gamma^{(WW\gamma)}_{\mu\nu\rho}(p_1,p_2,p_3) 
= ie\left[\Theta^{\rm (SM)}_{\mu\nu\rho}(p_1,p_2,p_3)
+ \Delta\kappa_\gamma \Theta^{(\Delta\kappa)}_{\mu\nu\rho}(p_1,p_2,p_3) 
+ \frac{\lambda_\gamma}{M_W^2} \Theta^{(\lambda)}_{\mu\nu\rho}(p_1,p_2,p_3) 
\right]
\end{equation}
where the $\Theta$ tensors are, respectively,
\begin{eqnarray}
\Theta^{\rm (SM)}_{\mu\nu\rho}(p_1,p_2,p_3) &=& 
  g_{\mu\nu}  \left( p_1 - p_2  \right)_\rho 
+ g_{\nu\rho} \left( p_2 - p_3  \right)_\mu  
+ g_{\rho\mu} \left( p_3 - p_1  \right)_\nu  
 \\
\Theta^{(\Delta\kappa)}_{\mu\nu\rho}(p_1,p_2,p_3) &=& 
g_{\mu\rho} p_{3\nu} - g_{\nu\rho} p_{3\mu}  \nonumber \\
\Theta^{(\lambda)~~}_{\mu\nu\rho}(p_1,p_2,p_3) &=& 
p_{1\rho}p_{2\mu}p_{3\nu} - p_{1\nu}p_{2\rho}p_{3\mu}
- g_{\mu\nu} \left(p_{1\rho}p_2\cdot p_3 - p_{2\rho}p_3\cdot p_1\right) 
\nonumber \\
& - & g_{\nu\rho} \left(p_{2\mu}p_3\cdot p_1 - p_{3\mu}p_1\cdot p_2\right) 
- g_{\mu\rho} \left(p_{3\nu}p_1\cdot p_2 - p_{1\nu}p_2\cdot p_3\right) 
\nonumber
\end{eqnarray}
This is the most general form consistent with the gauge and Lorentz 
symmetries of the SM\cite{Dubinin}. The extra terms whose coefficients 
are $\Delta\kappa_ \gamma$ and $\lambda_\gamma$ respectively are known as 
the {\it anomalous} TGV's. Noting that the terms in 
$\Theta^{(\Delta\kappa)}_{\mu\nu\rho}$ also appear in $\Theta^{\rm 
(SM)}_{\mu\nu\rho}$, one can also combine the terms and use 
$\kappa_\gamma = 1 + \Delta\kappa_\gamma$, but in this paper we have used 
only $\Delta\kappa_\gamma$, which agrees with the common usage by most 
experimental collaborations.

These anomalous TGV's have been studied in some detail in many processes, 
both at low energies and at high energies \cite{Barklow}. No evidence for 
any deviation from the SM has been found till date, as a result of which, 
we have fairly stringent upper bounds on the anomalous couplings 
$\Delta\kappa_\gamma$ and $\lambda_\gamma$. The strongest bounds come 
from the study of $W^+W^-$ production at the Large Electron Positron 
(LEP) collider at CERN, Geneva \cite{LEP}. The early runs of the LHC have 
also yielded bounds published by both the ATLAS and the CMS 
Collaborations \cite{ATLAS,CMS}, but these are not, as yet, competitive 
with the LEP bounds. A summary of the best available constraints on 
$\Delta\kappa_\gamma$ and $\lambda_\gamma$ is given in 
Table~\ref{tab:bounds}.

\begin{table}[h!]
\begin{center}
\begin{tabular}{cccccc}
                      & LEP\cite{LEP}   & CDF\cite{CDF}    & D0\cite{D0}    
                      & ATLAS\cite{ATLAS} & CMS\cite{CMS} \\
\hline\hline
$\Delta\kappa_\gamma$ & [-0.099, 0.066] & [-0.460, 0.390]  & [-0.158, 0.255]      
                      & [-0.135, 0.190] & [-0.210, 0.220] \\
$\lambda_\gamma$      & [-0.059, 0.017] & [-0.180, 0.170]  & [-0.036, 0.044]      
                      & [-0.065, 0.061] & [-0.048, 0.037] \\
\hline
\end{tabular}
\caption{\footnotesize Allowed ranges, at 95\% C.L., on the anomalous 
$WW\gamma$ couplings from the data collected at the LEP, Tevatron and LHC 
experiments. In each case, the most restrictive of the reported 
measurements is taken.}
\label{tab:bounds}
\end{center}
\end{table}
\vspace*{-0.2in}
Although these constraints -- especially the ones from the LEP data -- 
are fairly stringent, they come with some caveats, viz. the fact that the 
processes used to put these bounds on the $WW\gamma$ anomalous TGV's are 
often affected by the $WWZ$ anomalous TGV's. For example, if we consider 
the LEP process $e^+e^- \to W^+W^-$ through an $s$-channel photon 
exchange, there is also a similar process through an $s$-channel $Z^0$ 
exchange. The bounds quoted in Table~\ref{tab:bounds} are sometimes 
obtained with the assumption that there are anomalous couplings in the 
$WW\gamma$ vertex alone, but not in the $WWZ$ vertex, and sometimes by 
assuming both kinds of anomalous couplings exist and may or may not be 
equal. Moreover, since these anomalous couplings lead to unitarity 
violation at high energies, sometimes they are taken with arbitrary 
factors of the form $(1 + s/\Lambda^2)^\alpha$, where $\Lambda$ is a high 
energy scale, and $\alpha$ is an adjustable exponent\cite{CDF}. Not every 
experimental collaboration, however, uses these factors, and hence 
comparison of the different constraints could be deceptive. Further, 
there always remains a possibility that there may be anomalous couplings 
in both $WW\gamma$ and $WWZ$ vertices such that these interfere 
destructively to produce a very small effect. In such a situation, many 
of the above bounds could be rendered invalid. A cleaner mode is the 
study of $W\gamma$ (or $WZ$) final states at a hadron collider, but this 
suffers from the problem of low cross sections and large SM backgrounds.  
Photoproduction of $W$ and $Z$ bosons have also been studied in the 
context of $ep$ colliders like the DESY HERA \cite{HERA} and the proposed 
CERN LHeC \cite{Brazil}, but these do not probe very small values of the 
anomalous TGV couplings, and moreover, $\gamma^\ast \to WW$ production 
can easily get mixed with $Z^\ast \to WW$ processes.

In this context, we wish to point out that at an $ep$ collider one can 
clearly distinguish between charged current (CC) events $e + p \to \nu_e 
+ {\rm jet}$ arising from $W$ boson exchange, and neutral current (NC) 
events $e + p \to e + {\rm jet}$ arising from photon or $Z$ boson 
exchange, simply by triggering on the missing energy or the electron in 
the final state. Considering the CC events, if a photon is radiated from 
the exchanged $W$ boson, we can trigger on a final state with a photon, 
one (or more) jets and missing energy. The crucial point to note is that 
if we trigger on a final state photon, there will be no interference from 
the $WWZ$ vertex, anomalous or otherwise. Thus, if we trigger on a final 
state photon, an $ep$ collider can provide very clean bounds on the 
anomalous TGV's and this is what is investigated in the present work.

\begin{figure}[h!] 
\centerline{\epsfig{file=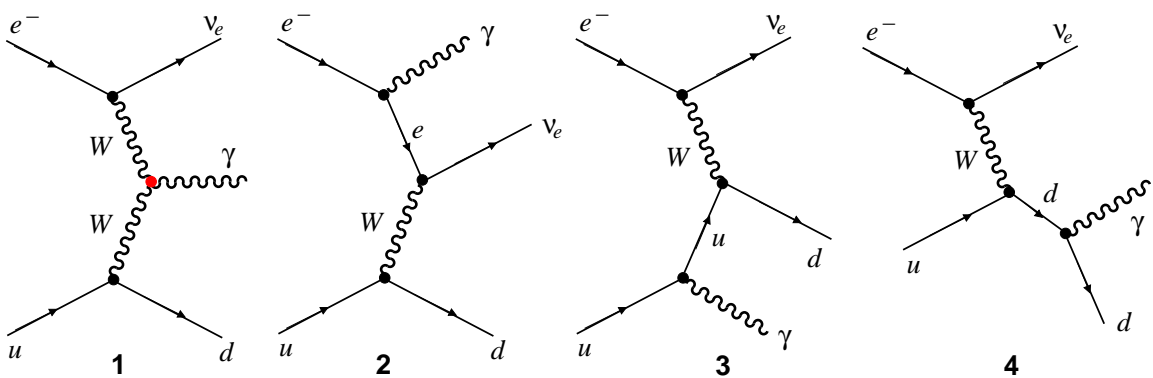,scale = 0.35}}
\caption{\footnotesize Feynman diagrams contributing at parton level to 
the process $e^- p \to \nu_e + \gamma + {\rm jet}$. The red dot in the 
diagram marked '1' corresponds to the anomalous TGV.}
\label{fig:Feynman}
\end{figure}

The possible diagrams which give rise to the process $e + p \to \nu_e + 
{\rm jet}$ in the framework of the SM are given in 
Fig.~\ref{fig:Feynman}. The graph marked `1' has a red dot indicating the 
contribution of possible anomalous $WW\gamma$ coupling terms.

Evaluation of the diagrams in Fig.~\ref{fig:Feynman} leads to a matrix 
element of the form
\begin{equation}
{\cal M} = {\cal M}_0 + \Delta\kappa_\gamma {\cal M}_1 
+ \lambda_\gamma {\cal M}_2
\end{equation}
where the dominant term ${\cal M}_0$ is the Standard Model contribution, 
which arises from all four diagrams, and the trailing terms 
$\Delta\kappa_\gamma {\cal M}_1$ and $\lambda_\gamma {\cal M}_1$ get 
contributions only from the diagram marked `1'. Squaring and 
spin-summing/averaging this matrix element and integrating it over the 
accessible phase space leads, then to a parton-level cross-section of the 
generic form
\begin{equation}
\hat{\sigma} =  \hat{\sigma}_{00} + \Delta\kappa^2_\gamma \hat{\sigma}_{11}
+ \lambda^2_\gamma \hat{\sigma}_{22} + \Delta\kappa_\gamma \hat{\sigma}_{01} 
+\lambda_\gamma \hat{\sigma}_{02} 
+ \Delta\kappa_\gamma\lambda_\gamma \hat{\sigma}_{12}
\end{equation}
where, in general, $\hat{\sigma}_{ij}$ arises from integration of terms 
of the form $\sum_s {\cal M}_i^\dagger {\cal M}_j$. Given the small 
values of $\Delta\kappa_\gamma$ and $\lambda_\gamma$ allowed by the 
experimental data (see Table~\ref{tab:bounds}), it is clear that the 
dominant new physics contributions will come from the interference terms 
$\Delta\kappa_\gamma \hat{\sigma}_{01}$ and $\lambda_\gamma 
\hat{\sigma}_{02}$, which vary linearly with the anomalous coupling 
parameters $\Delta\kappa_\gamma$ and $\lambda_\gamma$.Thus, the main 
question is whether these terms can be at all significant when compared 
to the dominant SM term $\hat{\sigma}_{00}$.

In the most general case, the answer to the above question is clearly in 
the negative, since similar considerations will hold at any collider, 
including the CERN LEP, Fermilab Tevatron and CERN LHC, all of which have 
already come up with negative results for their searches (see Table 
\ref{tab:bounds}). In fact, even if we take a differential cross-section 
for some kinematic variable $\xi$, we would still have
\begin{equation}
\frac{d\hat{\sigma}}{d\xi} =  \left(\frac{d\hat{\sigma}}{d\xi}\right)_{00} 
+ \Delta\kappa^2_\gamma \left(\frac{d\hat{\sigma}}{d\xi}\right)_{11}
+ \lambda^2_\gamma \left(\frac{d\hat{\sigma}}{d\xi}\right)_{22} 
+ \Delta\kappa_\gamma \left(\frac{d\hat{\sigma}}{d\xi}\right)_{01} 
+\lambda_\gamma \left(\frac{d\hat{\sigma}}{d\xi}\right)_{02} 
+ \Delta\kappa_\gamma\lambda_\gamma 
\left(\frac{d\hat{\sigma}}{d\xi}\right)_{12}
\label{eqn:diffcs}
\end{equation}
and any deviation of the observed deviation from the SM term (marked 00) 
would have been detected in the previous runs of these machines. The 
question is, therefore, if, in the context of the LHeC, we can ($a$) find 
some suitable variable $\xi$ which will show appreciable deviation 
between the left side of Eqn.~\ref{eqn:diffcs} and the SM term on the 
right, and ($b$) if we can devise a suitable set of kinematic cuts which 
will reduce the SM contribution as much as possible without affecting the 
$\Delta\kappa$ and $\lambda$ terms too much. In pursuit of the first 
goal, we require to go beyond the usual transverse momentum and 
pseudorapidity variables and use, instead, an azimuthal angle variable, 
which has been used quite successfully in the literature to predict 
detection techniques for anomalous $HWW$ couplings \cite{HWW}.
 
We have, of course, studied a fairly comprehensive set of the different 
possible kinematic variables that can be constructed using the final 
state particles. The one which we find most sensitive to the anomalous 
TGV's, especially the $\Delta\kappa_\gamma$ variable, is constructed as 
follows. The final state consists of an isolated hard photon and a single 
jet, with a substantial amount of missing transverse momentum. We now 
consider the transverse momenta of the jet ($\vec{p}_T^{\rm J}$) and the 
missing transverse momentum ($\not{\!\vec{p}}_T$) as two-dimensional 
vectors in the plane perpendicular to the beam axis. The angle between 
these vectors will be denoted by $\Delta\phi(J\!\!\not{\!p}_T)$ and can 
be constructed from
\begin{equation}
\cos \Delta\phi(J\!\!\not{\!p}_T) = 
\frac{\vec{p}_T^{\rm J} \cdot \not{\!\vec{p}}_T}
{\mid\vec{p}_T^{\rm J}\mid \, \mid\not{\!\vec{p}}_T\mid}
\end{equation}
We then set the variable $\xi = \Delta\phi(J\!\!\not{\!p}_T)$ in 
Eqn.~\ref{eqn:diffcs} for the rest of our analysis. The other kinematic 
variables where there are differences in contribution between the SM and 
other terms play their part in the following kinematic cuts.
\vspace*{-0.2in}
\begin{enumerate}
\item The emitted photon should have $p_T^\gamma \geq 50$~GeV.
\vspace*{-0.1in}
\item At least one final state jet should have $p_T^{\rm J} \geq 20$~GeV.
\vspace*{-0.1in}
\item The missing transverse momentum must satisfy $\not{\!\vec{p}}_T
\geq 20$~GeV.
\vspace*{-0.1in}
\item The pseudorapidities of the photon and the jet must satisfy
$\eta_\gamma, \eta_{\rm J} \leq 3.5$.
\vspace*{-0.1in}
\item The photon must be isolated from the jet by the criterion
$\Delta R_{\gamma{\rm J}} \geq 1.5$.   
\end{enumerate}
\vspace*{-0.2in}
Of these cuts, only the first and the last ones really function as 
selection cuts, since the others are practically forced upon us by the 
acceptance criteria of any standard detector which may be used at the 
LHeC\cite{LHeCTDR}. However, these two cuts, which together enforce the 
requirement of a hard, isolated photon, are most crucial in suppressing 
the SM background, the bulk of which obviously comes from bremsstrahlung 
processes, with their usual soft and collinear dominance.

It may be noted, at this stage, that these kinematic cuts have been 
specifically chosen to favour the $\Delta\kappa_\gamma$ terms in the 
cross-section. This is because the bounds on the $\lambda_\gamma$ 
coupling are already pretty strong, whereas the bounds on 
$\Delta\kappa_\gamma$ are much weaker. We, therefore, seek to maximise 
sensitivity to the $\Delta\kappa_\gamma$ variable. It may be pointed out, 
that as far as kinematic cuts go, the present ones are rather moderate. 
Far more stringent cuts are commonly used in LHC studies --- for example, 
in searches for supersymmetry signals one often encounters a demand that 
$\not{\!\vec{p}}_T \geq 300$~GeV\cite{LHCSUSY}. However, we have kept the 
cuts very conservative in this analysis, for two reasons. In the first 
place, more stringent cuts may end up by removing the entire signal, 
given that the LHeC integrated luminosity may not be all that high. 
Moreover, we find that stronger cuts on the other variables tend, in 
general, to reduce the difference between the SM and the anomalous TGV 
terms in the $\Delta\phi(J\!\!\not{\!p}_T)$ distribution. Thus, the above 
cuts are essentially chosen to maximise this difference, insofar as a 
rounding-off to standard values allows.

\begin{figure}[h!] 
\centerline{\epsfig{file=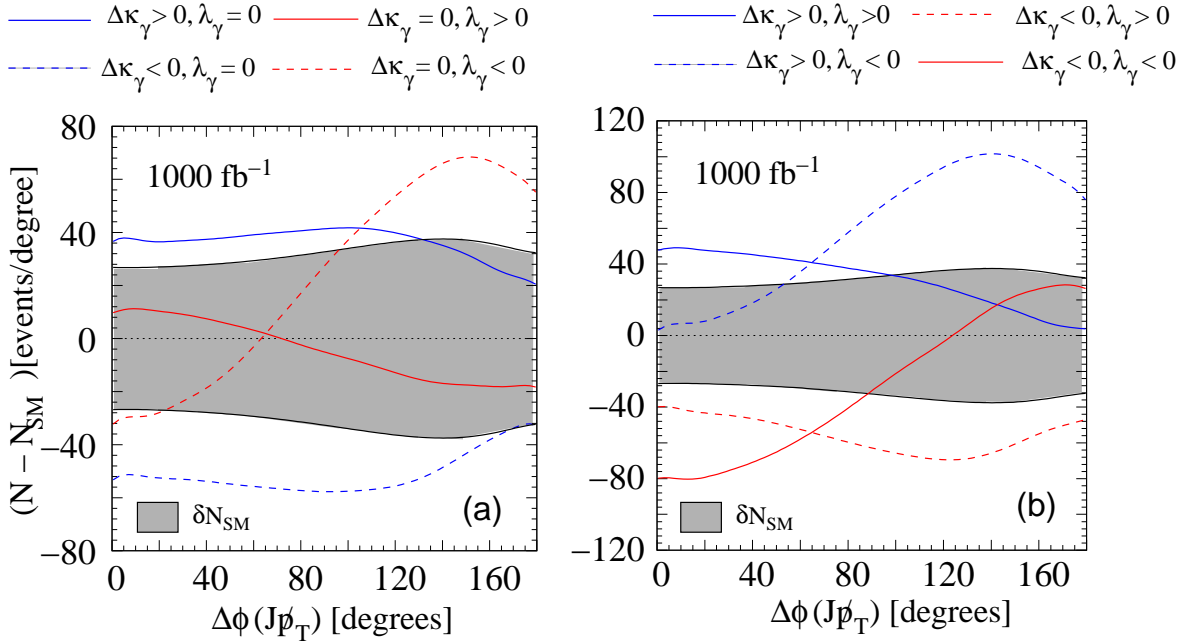,scale = 0.5}}
\caption{\footnotesize Distribution of the final events in the azimuthal 
angle difference variable $\Delta\phi ({\rm J}\!\not{\!\!p}_T)$. In each 
graph on the panel on the left, marked ($a$), one of the parameters 
$\Delta\kappa_\gamma$ or $\lambda_\gamma$ is set to zero, while the other 
is set to the maximum or the minimum value allowed by LEP data at the 
95\% C.L. level. The panel on the right, marked ($b$) shows similar 
distributions, but for the cases when both the variables have non-zero 
values corresponding to the maximum or minimum value allowed by LEP data 
at the 95\% C.L. level. In both panels, the shaded region shows the 95\% 
C.L. fluctuation in the SM background. To generate this figure, we have 
assumed an integrated luminosity of $10^3$~fb$^{-1}$ and an electron beam 
energy of 140~GeV.}
\label{fig:azimuthal}
\end{figure}

Some of our results are plotted in Fig.~\ref{fig:azimuthal} for the case 
of a 140~GeV electron beam colliding with the 7~TeV protons from the LHC. 
On both the left and right panels, we have plotted the variable 
$\Delta\phi(J\!\!\not{\!p}_T)$ (in degrees) on the abscissa, while the 
ordinate represents the difference in number of events between the signal 
and the background. These numbers are obtained from the differential 
cross-section multiplied by an integrated luminosity estimated at 
$10^3$~fb$^{-1}$.

In both panels of Fig.~\ref{fig:azimuthal}, the region shaded grey 
indicates the 95\% C.L. fluctuation $\delta N$ in the SM background. This 
is essentially flat, apart from a minor increase on the right side of the 
plot. The left panel, marked (a), shows the distribution when one of the 
couplings $\Delta\kappa_\gamma$ or $\lambda_\gamma$ is at its maximum 
magnitude as permitted by the LEP constraints at 95\% C.L. Thus, the 
solid (dashed) blue lines indicate the cases when $\lambda_\gamma = 0$ 
and $\Delta\kappa_\gamma = 0.066\, (-0.099)$. Similarly the solid 
(dashed) red lines indicate the cases when $\Delta\kappa_\gamma = 0$ and 
$\lambda_\gamma = 0.017\, (-0.059)$. It is clear that at 95\% C.L. it 
will be possible to distinguish the anomalous TGV effects from the SM 
background for both extreme values of $\Delta\kappa_\gamma$, as well as 
some intermediate values which are not too small in magnitude. On the 
other hand, the deviation from the SM is too small for positive values of 
$\lambda_\gamma$, though when $\lambda_\gamma$ goes negative, some 
deviation would be observed in the backward direction. In general, of 
course, both $\Delta\kappa_\gamma$ and $\lambda_\gamma$ could have 
nonzero values, and the corresponding effects are illustrated in the 
right panel of Fig.~\ref{fig:azimuthal}, marked (b). Here the solid red 
(blue) lines indicate the cases when both $\Delta\kappa_\gamma$ and 
$\lambda_\gamma$ have their maximum allowed negative (positive) values, 
as given above. In both cases there are deviations in the first quadrant, 
with the negative values leading to somewhat greater deviations than the 
positive ones.

At this point it is necessary to remember that the curves in panel (b) 
are not just the superposition of the relevant curves in panel (a) 
because of the interference term proportional to $\hat{\sigma}_{12}$ in 
Eqn.~\ref{eqn:diffcs}. When $\Delta\kappa_\gamma$ and $\lambda_\gamma$ 
have opposing signs, we get greater deviations from the background, with 
the case $\Delta\kappa_\gamma < 0$ yielding deviations in the first 
quadrant, and the case $\Delta\kappa_\gamma > 0$ yielding deviations in 
the second quadrant. Of course, the curves shown in 
Fig.~\ref{fig:azimuthal} represent only the extreme values of the 
anomalous couplings, as well as a highly optimistic estimate of the 
integrated luminosity. We thus need to set up a more sensitive criterion 
for distinguishability than mere inspection of a graph like 
Fig.~\ref{fig:azimuthal}. In order to do this, we divide the range of 
$\Delta\phi ({\rm J}\!\not{\!\!p}_T)$ into 36 bins, i.e. of $5^0$ each, 
and then calculate a $\chi^2(\Delta\kappa_\gamma, \lambda_\gamma)$ as 
follows
\begin{equation}
\chi^2(\Delta\kappa_\gamma, \lambda_\gamma) = \sum_{i=1}^{36} 
\frac{\left[N_i^{\rm tot}(\Delta\kappa_\gamma, \lambda_\gamma) - N_i^{\rm 
SM}\right]^2}{\left[\delta{N_i^{\rm SM}}\right]^2}
\label{eqn:chisq}
\end{equation}     
where $N_i^{\rm SM} = N_i^{\rm tot}(\Delta\kappa_\gamma = 0, 
\lambda_\gamma = 0)$ and, if the numbers are large enough, 
$\delta{N_i^{\rm SM}} = \sqrt{N_i^{\rm SM}}$. Since $N_i^{\rm 
tot}(\Delta\kappa_\gamma, \lambda_\gamma) = {\cal L} \sigma_i^{\rm 
tot}(\Delta\kappa_\gamma, \lambda_\gamma)$, where ${\cal L}$ is the 
integrated luminosity and $\sigma_i^{\rm tot}(\Delta\kappa_\gamma, 
\lambda_ \gamma)$ is the cross-section in that bin, it follows that 
$\chi^2(\Delta\kappa_\gamma, \lambda_\gamma)$ varies linearly with ${\cal 
L}$. If the difference between the observed difference and the SM 
prediction arises from random fluctuations, we should obtain a value of 
$\chi^2 \simeq 23.268$ at 95\% C.L.. The criterion for a 95\% C.L. 
discovery, then, is simply
\begin{equation}
\chi^2(\Delta\kappa_\gamma, \lambda_\gamma) > 23.268 \ .
\label{eqn:discovery}
\end{equation} 
The usefulness of this criterion is illustrated below, in Fig.~ 
\ref{fig:kappareach}. Here we have set $\lambda_\gamma = 0$ and plotted, 
as a function of the integrated luminosity ${\cal L}$, the minimum value 
of $\Delta\kappa_\gamma$ for which the criterion in 
Eqn.~\ref{eqn:discovery} is satisfied, i.e. the anomalous coupling 
$\Delta\kappa_\gamma$ is discoverable at the LHeC.

\begin{figure}[h!] 
\begin{minipage}[h]{0.57\linewidth}
\centerline{\epsfig{file=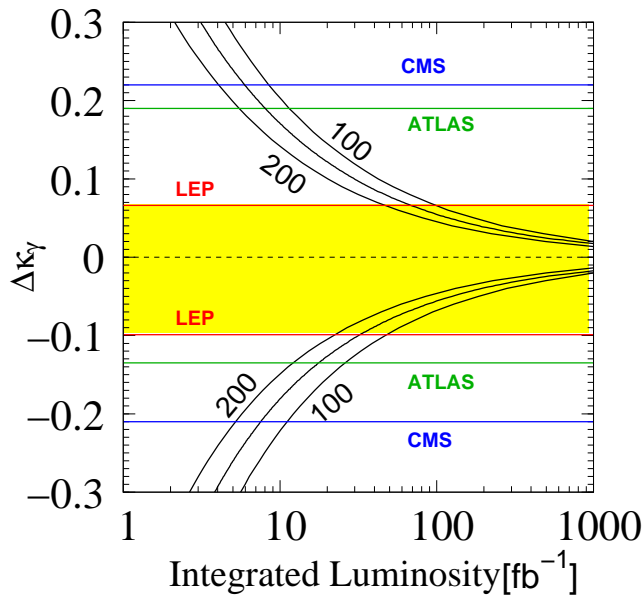,scale = 0.54}}
\vspace*{-0.1in}
\caption{\footnotesize The 95\% C.L. discovery reach of the LHeC in 
$\Delta\kappa_\gamma$ (assuming $\lambda_\gamma = 0$) as a function of 
the integrated luminosity. The best experimental bounds at 95\% C.L. are 
as indicated on the graph. The three solid lines correspond respectively 
to electron beam energies of 100~GeV (marked), 140~GeV and 200~GeV 
(marked).}
\label{fig:kappareach}
\end{minipage} 
\hspace{0.02\linewidth}
\begin{minipage}[h]{0.41\linewidth}
In Fig.~\ref{fig:kappareach}, the three solid curves represent three 
possible energies, viz. 100~GeV, 140~GeV and 200~GeV of the electrons 
colliding with 7~TeV protons. As may be expected from the high momentum 
dependence of the anomalous couplings, we get somewhat better results 
with higher energy electrons than with lower energy electrons, though the 
difference is not all that important. On the other hand, increase of 
luminosity allows us to probe smaller and smaller values of 
$|\Delta\kappa_\gamma|$, as is apparent from the the converging lines in 
the figure. For comparison, we have also plotted the constraints on 
$\Delta\kappa_\gamma$ from the ATLAS and CMS experiments, as well as the 
combined LEP collaborations. The LEP bounds, which are the most 
restrictive, are highlighted in yellow to make the comparison easy.  
\end{minipage} 
\end{figure} 
It may be immediately noted that, as of now, only the LEP bounds will be 
comparable with the LHeC results, using the azimuthal angle variable, as 
soon as the integrated luminosity crosses a few tens of fb$^{-1}$. 
However, in order to better the LEP results, we will require an 
integrated luminosity of about 50, 70 or 100~fb$^{-1}$ for 
$\Delta\kappa_\gamma >0$ for an electron beam energy of 200, 140 or 
100~GeV respectively. For $\Delta\kappa_\gamma <0$, the corresponding 
values are about 25, 30 and 50~GeV respectively. Thus, we may conclude 
that an integrated luminosity of 100~fb$^{-1}$, or more, will enable the 
LHeC to become the most powerful probe of the anomalous TGV 
$\Delta\kappa_\gamma$ till now.

\begin{figure}[h!] 
\begin{minipage}[h]{0.41\linewidth}
The solid (black) contours in Fig.~\ref{fig:contourplot} represent the 
discovery reach of the LHeC, using the azimuthal angle difference 
variable $\Delta\phi ({\rm J}\!\not{\!\!p}_T)$ and the $\chi^2$ technique 
of Eqns.~\ref{eqn:chisq} and \ref{eqn:discovery}. In each case the 
integrated luminosity, in fb$^{-1}$, is marked alongside the relevent 
contour.Regions lying between the central point $\Delta\kappa_\gamma = 
\lambda_ \gamma = 0$ and each contour are {\it inaccessible} for that 
value of integrated luminosity. For this graph, we have assumed an 
electron beam energy of 140~GeV. Obviously the contours will shrink 
marginally if the electron energy is increased and vice versa. For 
comparison, we have also superposed on these contour plots the correlated 
95\% C.L. constraints from ($a$) the CDF and D0 Collaborations (dashes, 
green) at the Fermilab Tevatron, ($b$) the ATLAS and CMS constraints 
(dashes, blue) from the LHC, and ($c$) the LEP constraints (solid, red 
and shaded yellow).
\end{minipage} 
\hspace{0.02\linewidth}
\begin{minipage}[h]{0.57\linewidth}
\centerline{\epsfig{file=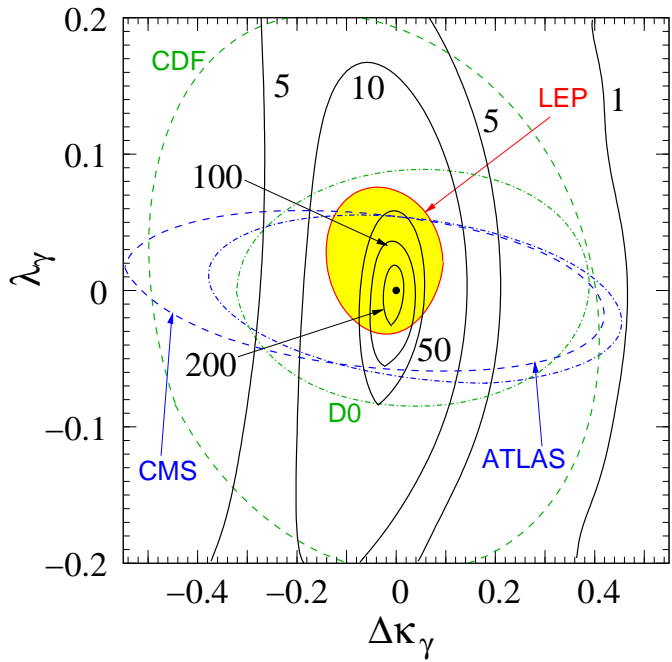,scale = 0.5}}
\vspace*{-0.1in}
\caption{\footnotesize 95\% C.L. discovery contours in the 
$\Delta\kappa_\gamma$--$\lambda_\gamma$ plane corresponding to an 
electron beam energy of 140~GeV. The dot in the centre represents the 
Standard Model value. The region between this dot and each contour is 
{\it not} discoverable for the luminosity (in fb$^{-1}$) marked alongside 
the contour. The different experimental bounds at 95\% C.L. are also 
exhibited.}
\label{fig:contourplot}
\end{minipage} 
\end{figure}

The graphs in Fig.~\ref{fig:kappareach} do not tell the whole story, 
however, for they represent the specific case when $\lambda_\gamma = 0$. 
In general, as we have seen in Fig.~\ref{fig:azimuthal}, the results will 
be different when both types of anomalous couplings assume non-zero 
values. We have, therefore, made a study of the joint variation of the 
$\chi^2$ variable in Eqn.~\ref{eqn:chisq} with both $\Delta\kappa_\gamma$ 
and $\lambda_\gamma$ varying over their allowed ranges. Our results are 
illustrated in Fig.~\ref{fig:contourplot}, where we have plotted 
discovery limits as contours in the plane of $\Delta\kappa_\gamma$ and 
$\lambda_\gamma$. Obviously, the black dot in the centre of the graph, 
which corresponds to $\Delta\kappa_\gamma = \lambda_\gamma = 0$, is the 
SM prediction. i.e. no anomalous TGV's.

It is immediately obvious from Fig.~\ref{fig:contourplot} that even with 
${\cal L} = 100$~fb$^{-1}$, the LHeC can already access part of the 
parameter space which was inaccessible to the LEP and has been hitherto 
inaccessible at hadron colliders as well. With ${\cal L} = 200$~fb$^{-1}$ 
it is apparent that the LHeC results will surpass all existing bounds, 
and it is easy to guess that the inaccessible region shrinks to really 
small values if the luminosity can be taken as high as ${\cal L} = 
1000$~fb$^{-1}$.

In this work, therefore, we have shown that the LHeC can provide a very 
powerful probe of the anomalous TGV's, if we use the azimuthal angle 
difference variable hitherto mainly proposed to study Higgs boson 
physics. It is still unknown how well these discovery limits will compare 
with the results of the LHC, when we consider its run at 13-14~TeV and an 
integrated luminosity of a thousand fb$^{-1}$ or more, for the same 
variable can be used to complement and enhance other studies proposed 
using the transverse momentum and other, more conventional distributions. 
Irrespective of that, however, the LHeC has the nice feature that one can 
pin down the $WW\gamma$ vertex separately, and hence, this result will 
have no contamination from possible anomalous effects in the $WWZ$ 
vertex. This, in itself, is a strong motivation to build and run the 
LHeC.

\footnotesize {\it Acknowledgements}: Some of this work was done during 
the 13th Workshop on High Energy Physics Phenomenology (WHEPP-13) at 
Puri, India, during December 2013. The work of SR is partly funded by the 
Board of Research in Nuclear Studies, Government of India, under project 
no. 2013/37C/37/BRNS.
 

\end{document}